\documentclass[12pt,,letterpaper]{JHEP3}
\usepackage{graphics}
\usepackage{epsfig}

\title{Refractive index in holographic superconductors}

\author{
Xin Gao\\
Key Laboratory of Frontiers in Theoretical Physics, Institute of
Theoretical Physics, Chinese Academy of Sciences, 100190 Beijing,
China\\
\email{xgao@itp.ac.cn}}

\author{Hongbao Zhang\\
Crete Center for Theoretical Physics, Department of Physics, \\
University of Crete, 71003 Heraklion, Greece\\
\email{hzhang@physics.uoc.gr}}

\abstract{With the probe limit, we investigate the behavior of the
electric permittivity and effective magnetic permeability and
related optical properties in the s-wave holographic
superconductors. In particular, our result shows that unlike the
strong coupled systems which admit a gravity dual of charged black
holes in the bulk, the electric permittivity and effective magnetic
permeability are unable to conspire to bring about the negative
Depine-Lakhtakia index at low frequencies, which implies that the
negative phase velocity does not appear in the holographic
superconductors under such a situation.}

\begin{document}
\section{Introduction}
Since the advent of AdS/CFT correspondence, not only do the main
efforts include the examination of its validity case by case, but
also involve its applications in various fields, in which condensed
matter physics has recently received a special attention. The
motives are twofold. On the one hand, there exists many strong
coupled systems in condensed matter physics, which are intractable
by the traditional approaches. While AdS/CFT correspondence, as kind
of strong/weak duality, can provide a powerful tool to attack them.
On the other hand, unlike other disciplines such as particle physics
and cosmology, condensed matter physics allows one to engineer
matter such that various vacuum states and phases can be created in
the laboratory, which may in turn provide the first experimental
evidence for AdS/CFT correspondence. For a review of this subject,
please refer to \cite{Hartnoll,Herzog1,McGreevy,Sachdev}.

 In particular, most recently such a bulk/boundary duality has been
applied to explore the optical properties of strong coupled media
which admit charged black holes as a dual gravitational description
and it is shown that this type of media generally have a negative
refractive index at low frequencies\cite{AFMP}. Along this line,
this paper is a first attempt to investigate how the refractive
index and related optical properties behave in the holographic
superconductors and to see whether the negative refractive index
shows up at low frequencies.

The gravity dual of superconductors was firstly established to model
the s-wave superconductors\cite{Gubser1,HHH}. Later both the p-wave
and d-wave holographic superconductors were also
realized\cite{Gubser2,GP,Chen,Herzog2,BHY,Benini}. Please refer to
\cite{Horowitz} and references therein for a review of various
properties related to such holographic superconductors.

In the next section, we shall recall the holographic model of s-wave
superconductors, where to make our life easier we would like to work
in the probe limit and focus on the special case in which the
complex scalar field is massless. After the setup, in Section
\ref{mainresult} we will present our numerical results for the
behavior of the relevant optical quantities in such holographic
superconductors. Conclusions and discussions will be addressed in
the end.
\section{Holographic model of superconductors}
\subsection{Abelian-Higgs model and holographic dictionary}
The bulk dual contents of a holographic superconductor with a s-wave
order parameter in 4+1 dimensions include the gravitational field,
the $U(1)$ gauge field, and the complex scalar field of mass $m$ and
charge $q$. The corresponding bulk action is given by\cite{Herzog2}
\begin{equation}
S_{bulk}=\int
d^5x\sqrt{-g}[\frac{1}{2\kappa^2}(R+\frac{12}{L^2})-\frac{1}{q^2}(\frac{1}{4}F_{ab}F^{ab}+|D_a\Phi|^2+m^2|\Phi|^2)],
\end{equation}
where $F\equiv dA$, and $D\equiv \nabla-iA$ where $\nabla$ is the
covariant derivative compatible with the metric. In what follows, we
will work in the probe limit, i.e., $\frac{\kappa}{qL}\rightarrow 0$
in which gravity decouples from the matter fields and the
backreaction from the matter fields can be ignored. Thus a solution
to Einstein equation is a black brane, i.e.,
\begin{equation}
ds^2=\frac{L^2}{u^2}[-f(u)dt^2+dx^2+dy^2+dz^2+\frac{du^2}{f(u)}],
\end{equation}
where $f(u)=1-(\frac{u}{u_h})^4$ with $u_h>0$ the horizon.
Obviously, in the limit $u\rightarrow 0$, such a solution
asymptotically becomes anti-de Sitter. Next around this background
the equations of motion for the matter field are given by
\begin{eqnarray}
0&=&D_aD^a\Phi-m^2\Phi=\frac{1}{\sqrt{-g}}(\partial_a-iA_a)[\sqrt{-g}g^{ab}(\partial_b-iA_b)\Phi]-m^2\Phi,\nonumber\\
0&=&\nabla_a
F^{ab}-i(\bar{\Phi}D^b\Phi-\Phi\bar{D}^b\bar{\Phi})=\frac{1}{\sqrt{-g}}\partial_a(\sqrt{-g}F^{ab})-i[\bar{\Phi}(\partial^b-iA^b)\Phi-\Phi(\partial^b+iA^b)\bar{\Phi}].\nonumber\\
\end{eqnarray}
By the holographic dictionary, the boundary value of the metric
$g_{\mu\nu}$ acts as the source for the energy momentum stress
tensor $T^{\mu\nu}$ on the boundary, the bulk gauge field $A_\mu$
evaluated at the boundary serves as the source for a conserved
current $J^\mu$ associated with a global $U(1)$ symmetry, and the
near boundary data of the scalar field $\Phi$ sources a scalar
operator $O$ with the conformal scaling dimension $\Delta$
satisfying $\Delta(\Delta-4)=m^2L^2$. In what follows, we shall be
focusing on the particular case, i.e., $m^2=0$, which yields an
operator of dimension four by normalizability. Whence the asymptotic
solution of $\Phi$ and $A_\mu$ can be expanded as
\begin{eqnarray}
\Phi&=&\frac{q}{\sqrt{L^3}}(\Phi^{(0)}+\Phi^{(1)}u^4+\cdot\cdot\cdot),\nonumber\\
A_\mu&=&A^{(0)}_\mu+\frac{q^2}{2L}[A^{(1)}_\mu
u^2-g^{(0)\rho\sigma}\partial_\rho(\partial_\sigma
A^{(0)}_\mu-\partial_\mu A^{(0)}_\sigma)
u^2\ln\frac{u}{\epsilon}]+\cdot\cdot\cdot \label{expansion}
\end{eqnarray}
with $\epsilon\ll 1$  a UV cutoff and
$g^0_{\mu\nu}=\lim_{u\rightarrow 0}\frac{u^2}{L^2}g_{\mu\nu}$. Then
it follows from the holographic dictionary that the expectation
value of the corresponding boundary quantum field theory operators
$\langle O\rangle$ and $\langle J^\mu\rangle$ can be obtained by
variations of the renormalized action with respect to the sources,
i.e.,
\begin{eqnarray}
\langle O\rangle&=&\lim_{u\rightarrow
0}\frac{1}{\sqrt{-g^0}}\frac{\delta
S}{\delta\bar{\Phi}^{(0)}}=\Phi^{(1)},\nonumber\\
\langle J^\mu\rangle&=&\lim_{u\rightarrow
0}\frac{1}{\sqrt{-g^0}}\frac{\delta S}{\delta
A^0_\mu}=g^{(0)\mu\nu}[A^{(1)}_\nu+cg^{(0)\rho\sigma}\partial_\rho(\partial_\sigma
A^{(0)}_\nu-\partial_\nu A^{(0)}_\sigma)],\label{expectation}
\end{eqnarray}
where the constant $c$ is renormalization scheme
dependent\cite{Skenderis}, to be fixed in the later discussions.
\subsection{Holographic normal phase and superconducting phase} With
the following ansatz, i.e.,
\begin{equation}
\Phi=\Phi(z),A_t=\phi(z),A_x=0,A_y=0,A_z=0,
\end{equation}
the equations of motion can be reduced to
\begin{eqnarray}
0&=&\Phi''+(\frac{f'}{f}-\frac{3}{u})\Phi'+\frac{\phi^2}{f^2}\Phi,\nonumber\\
0&=&\phi''-\frac{1}{u}\phi'-\frac{2L^2\Phi^2}{fu^2}\phi,\label{eomb}
\end{eqnarray}
where the prime denotes the differentiation with respect to $u$ and
 $\Phi$ has been taken to be real by taking into account the fact that
the $u$ component of Maxwell equations implies that the phase of
$\Phi$ is independent of $u$. Note that the regularization condition
requires $\phi=0$ on the horizon. Then multiplying the first
equation in (\ref{eomb}) by $f$, one can find $\Phi'=0$ on the
horizon. Thus for the above two second differential equations, we
have only a two parameter family of solutions, which can be labeled
by $\Phi(u_h)$ and $\phi'(u_h)$. For convenience, we will set $q=1$,
$L=1$, $u_h=1$ in the later calculations. The result for any other
$u_h$ can be easily restored by the following scaling rules, i.e.,
\begin{eqnarray}
\Phi^{(0)}(u_h)&=&\Phi^{(0)}(1),\nonumber\\
\Phi^{(1)}(u_h)&=&\frac{1}{u_h^4}\Phi^{(1)}(1),\nonumber\\
A^{(0)}_\mu(u_h)&=&\frac{1}{u_h}A^{(0)}_\mu(1),\nonumber\\
A^{(1)}_\mu(u_h)&=&\frac{1}{u_h^3}A^{(1)}_\mu(1),\nonumber\\
T(u_h)&=&\frac{1}{u_h\pi}.
\end{eqnarray}
\begin{figure}
\includegraphics[]{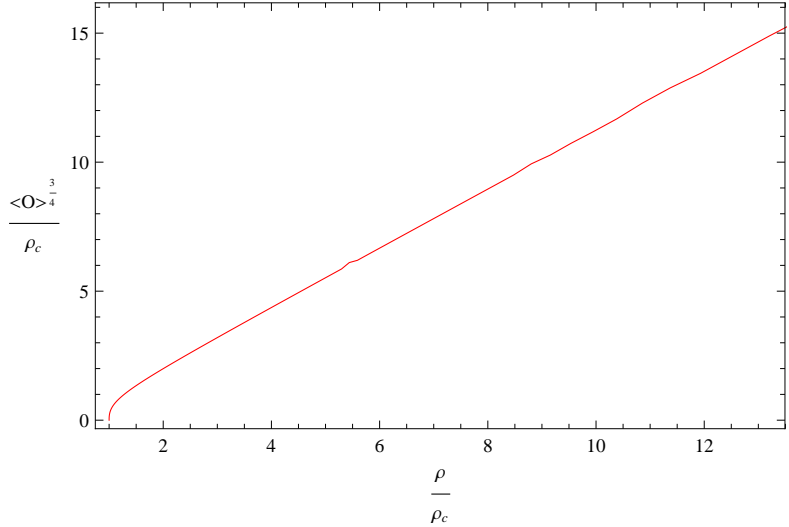}\\
  \caption{The condensate as a function of charge density, where the temperature is fixed to be $1$ as the reference scale.}\label{condensatevscharge}
\end{figure}
\begin{figure}
 \includegraphics[]{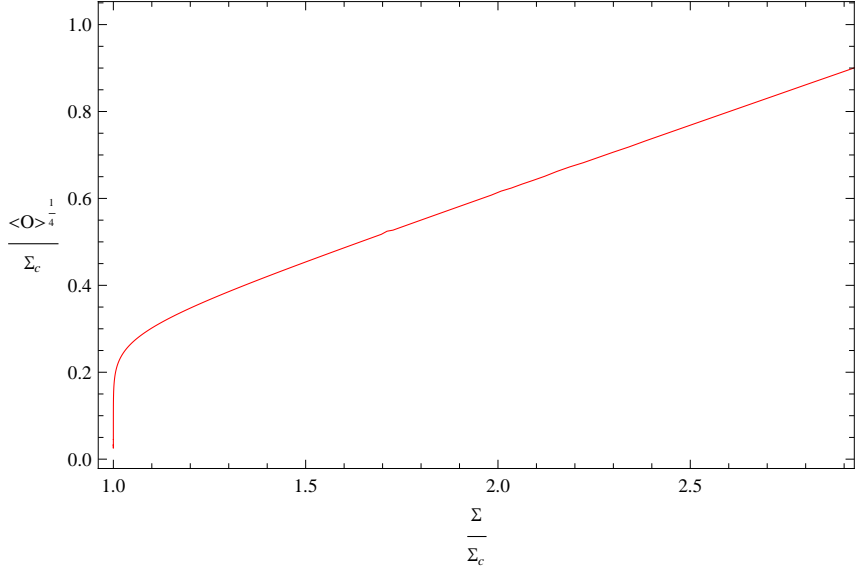}\\
  \caption{The condensate as a function of chemical potential, where the temperature is fixed to be $1$ as the reference scale.}\label{condensatevspotential}
\end{figure}

Firstly, it is easy to find the trivial solution as
\begin{equation}
\Phi=0,\phi=C(1-u^2),
\end{equation}
which corresponds to the normal phase on the boundary with the
chemical potential $C$. On the other hand, according to the
numerical calculations, by fixing the charge density or chemical
potential the non-trivial solution emerges below a critical
temperature $T_c$, and the condensate increases when the temperate
is lowered\cite{HR}. Its boundary correspondence is the
superconducting phase.

By the above scaling rules, such a non-trivial solution can also be
thought of as the superconducting phase emerging above a critical
charge density $\rho_c$ or a chemical potential $\Sigma_c$ if the
temperature is fixed. We would like to plot the corresponding
condensate as a function of charge density and chemical potential in
Fig.\ref{condensatevscharge} and Fig.\ref{condensatevspotential}
respectively.

\section{Refractive index in holographic superconductors\label{mainresult}}
\subsection{Holographic setup}
Before proceeding, we would like to introduce some quantities we
want to calculate and their relations without going into details. In
the linear response theory, the induced electromagnetic current $J$
is related to the external vector potential $A$ as
$J_i(\omega,k)=G_{ij}(\omega,k)A_j(\omega,k)$, where $G$ is the
retarded Green function. On the other hand, in the Laudau-Lifshitz
approach to electrodynamics of continuous media\cite{LL,MM,AG}, for
the isotropic media, the transverse part of the dielectric tensor is
determined by the transverse retarded Green function as
follows\footnote{Note that the expression in \cite{AFMP} is
different from ours due to the different conventions used for the
retarded Green function.}\cite{AFMP,DG}
\begin{equation}
\epsilon_T(\omega,k)=1+\frac{4\pi}{\omega^2}G_T(\omega,k),
\end{equation}
from which the corresponding electric permittivity and effective
magnetic permeability can be expressed as
\begin{eqnarray}
\epsilon(\omega)&=&1+\frac{4\pi}{\omega^2}G_{T0}(\omega),\nonumber\\
\mu(\omega)&=&\frac{1}{1-4\pi G_{T2}(\omega)}.
\end{eqnarray}
Here $G_{T0}(\omega)$ and $G_{T2}(\omega)$ are the expansion
coefficients of the transverse retarded Green function in $k$, i.e.,
\begin{equation}
G_T(\omega,k)=G_{T0}(\omega)+k^2G_{T2}(\omega)+\cdot\cdot\cdot
\end{equation}
with
$G_T(\omega,k)=(\delta^{ij}-\frac{k^ik^j}{k^2})G_{ij}(\omega,k)$.
With this, the refractive index can be given by
$n^2(\omega)=\epsilon(\omega)\mu(\omega)$. But to identify whether
our holographic superconductors display the opposite phase velocity
to the power flow, we shall appeal to the Depine-Lakhtakia index,
which is defined as
\begin{equation}
n_{DL}(\omega)=|\epsilon(\omega)|Re[\mu(\omega)]+|\mu(\omega)|Re[\epsilon(\omega)].
\end{equation}
As shown in \cite{DL}, the phase velocity is opposite to the power
flow if and only if $n_{DL}<0$.

 Now let us move on to the strategy to calculate these quantities by holography. By symmetry, in order to consider the refractive index in
the above holographic superconductors, it is enough to consider the
transverse fluctuations of the field $A$ as follows
\begin{equation}
\delta A_x=a_x(u)e^{-i\omega t+iky}.\label{perturbation}
\end{equation}
Note that such a fluctuation decouples and has no back reaction,
satisfying the equation of motion
\begin{equation}
0=a_x''+(\frac{f'}{f}-\frac{1}{u})a_x'+(\frac{\omega^2}{f^2}-\frac{k^2}{f}-\frac{2\Phi^2}{fu^2})a_x.\label{eomf}
\end{equation}
Whence $a_x$ goes like $(1-u)^{\pm i\frac{\omega}{4}}$ near the
horizon, corresponding to the outgoing and ingoing boundary
conditions respectively. We here choose the ingoing boundary
condition, which will lead to the retarded Green function in the
dual field theory\cite{SS}. Speaking specifically, with such an
ingoing boundary condition, we assume the corresponding asymptotic
expansion $a_x$ goes like
\begin{equation}
a_x=a^{(0)}_x+\frac{1}{2}[a^{(1)}_xu^2-(\omega^2-k^2)a^{(0)}_xu^2\ln\frac{u}{\epsilon}]+\cdot\cdot\cdot
\label{expansion}
\end{equation}
where we have used Eq.(\ref{perturbation}) and the second equation
in (\ref{expansion}). By Eq.(\ref{expectation}), the induced current
is given by
\begin{equation}
J_x=a^{(1)}_x+c(\omega^2-k^2)a^{(0)}_x.
\end{equation}
Whence the retarded transverse Green function can be obtained as
\begin{equation}
G_T(\omega,k)=\frac{a^{(1)}}{a^{(0)}}+c(\omega^2-k^2).
\end{equation}

An analytic solution of Eq.(\ref{eomf}) does not appear to be
available. So in the subsequent section, we shall resort to
Mathematica for numerical calculation, where in order to obtain
$G_{T0}(\omega)$ and $G_{T2}(\omega)$, it is advisable to adopt an
alternative approach by firstly expanding $a_x$ in series of $k$,
i.e.,
\begin{equation}
a_x=a_{x0}+k^2a_{x2}+\cdot\cdot\cdot,
\end{equation}
from which Eq.(\ref{eomf}) becomes
\begin{eqnarray}
0&=&a_{x0}''+(\frac{f'}{f}-\frac{1}{u})a_{x0}'+(\frac{\omega^2}{f^2}-\frac{2\Phi^2}{fu^2})a_{x0},\nonumber\\
0&=&a_{x2}''+(\frac{f'}{f}-\frac{1}{u})a_{x2}'+(\frac{\omega^2}{f^2}-\frac{2\Phi^2}{fu^2})a_{x2}-\frac{1}{f}a_{x0}.
\end{eqnarray}
Near the boundary, the asymptotic expansion for $a_{x0}$ and
$a_{x2}$ can be read out of Eq.(\ref{expansion}) as
\begin{eqnarray}
a_{x0}&=&a^{(0)}_{x0}+\frac{1}{2}(a^{(1)}_{x0}u^2-\omega^2a^{(0)}_{x0}u^2\ln\frac{u}{\epsilon})+\cdot\cdot\cdot\nonumber\\
a_{x2}&=&a^{(0)}_{x2}+\frac{1}{2}[a^{(1)}_{x2}u^2-(\omega^2a^{(0)}_{x2}-a^{(0)}_{x0})u^2\ln\frac{u}{\epsilon}]+\cdot\cdot\cdot.
\end{eqnarray}
Whence we have
\begin{eqnarray}
G_{T0}(\omega)&=&\frac{a^{(1)}_{x0}}{a^{(0)}_{x0}}+c\omega^2,\nonumber\\
G_{T2}(\omega)&=&\frac{a^{(1)}_{x0}}{a^{(0)}_{x0}}[\frac{a^{(1)}_{x2}}{a^{(1)}_{x0}}-\frac{a^{(0)}_{x2}}{a^{(0)}_{x0}}]-c,
\end{eqnarray}
where $c$ can be fixed by requiring that $G_{T0}$ approaches zero at
large frequencies as the system has no time to respond to the rapid
variation of external fields.
\subsection{Numerical results}
\begin{figure}

  \includegraphics[width=8cm]{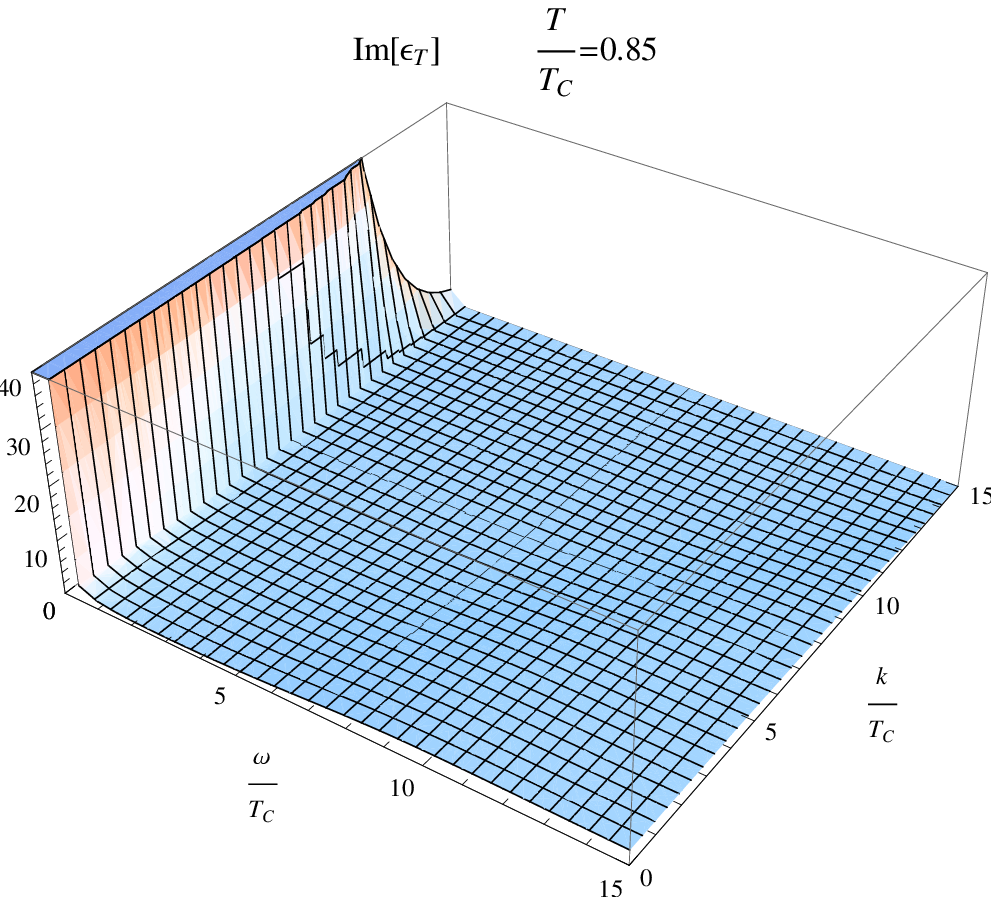}\\
  \caption{The imaginary part of the $\epsilon_T$ for real $\omega$ and $k$ at temperature $\frac{T}{T_c}=0.85$.}\label{3D1}
\end{figure}
\begin{figure}

  \includegraphics[width=8cm]{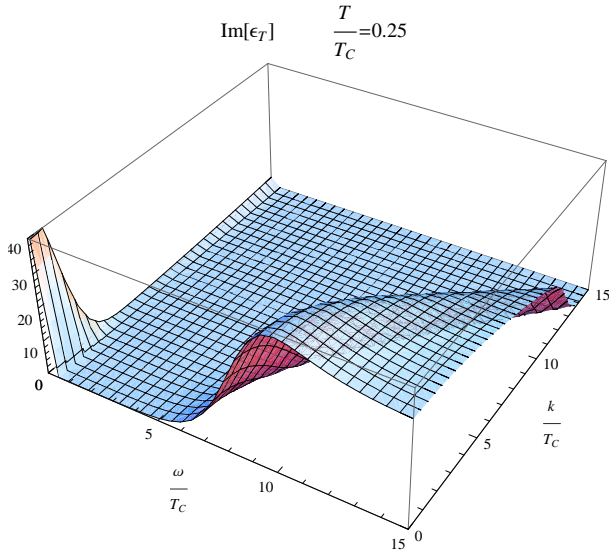}\\
  \caption{The imaginary part of the $\epsilon_T$ for real $\omega$ and $k$ at temperature $\frac{T}{T_c}=0.25$.}\label{3D4}
\end{figure}
Here we plot all the results by fixing the charge density to be $1$
but varying the temperature with respect to the critical temperature
and consequently the condensate.

As explained in \cite{AFMP}, firstly we have to ensure that the
system is in thermodynamical equilibrium, i.e., the imaginary part
of $G_T$, or equivalently the imaginary part of $\epsilon_T$ should
be greater than zero for real $\omega$ and $k$. This is actually
guaranteed by the holographic duality, as the bulk background is
static with the black brane. For illustration, we would like to plot
the imaginary part of $\epsilon_T$ in Fig.\ref{3D1} and
Fig.\ref{3D4} to demonstrate that this is the case.
\begin{figure}

  \includegraphics[width=10cm]{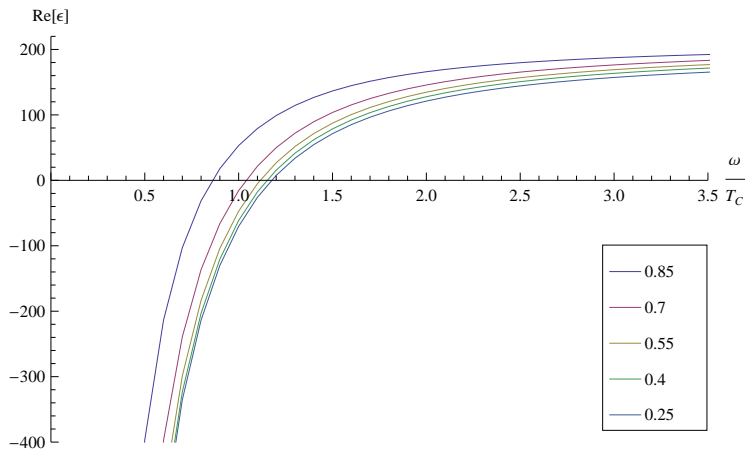}\\
  \caption{The real part of the electric permittivity as a function of frequency with various temperatures.}\label{permittivityreal}
\end{figure}
\begin{figure}

  \includegraphics[width=10cm]{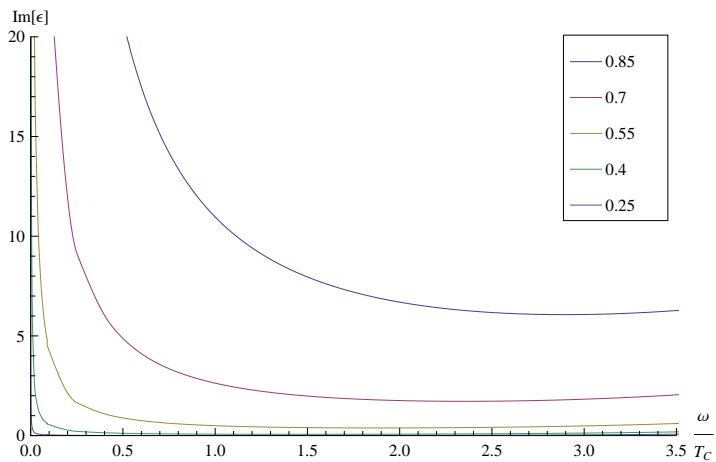}\\
  \caption{The imaginary part of the electric permittivity as a function of frequency with various temperatures.}\label{permittivityimaginary}
\end{figure}

Then we plot the real and imaginary parts of the electric
permittivity in Fig.\ref{permittivityreal} and
Fig.\ref{permittivityimaginary} respectively. By the scaling rules,
to lower the temperature is equivalent to increase the charge
density and chemical potential. Therefore the electric permittivity
here demonstrates the similar behavior at low frequencies as that
considered in \cite{AFMP}. In particular, the larger are the charge
density and chemical potential, the larger is the regime for the
negative real part of the electric permittivity. In addition, below
the critical temperature, both of the real and imaginary parts of
the electric permittivity diverge at $\omega=0$. By the fact that
the electric permittivity is related to the conductivity as
$\epsilon(\omega)=1+i\frac{4\pi}{\omega}\sigma(\omega)$, the real
part of the electric permittivity diverges as $-\frac{1}{\omega^2}$,
and accordingly the imaginary part blows up like
$\frac{\delta(\omega)}{\omega}$ due to the Kramers-Kronig relations
from causality\cite{Horowitz}.

\begin{figure}

  \includegraphics[width=10cm]{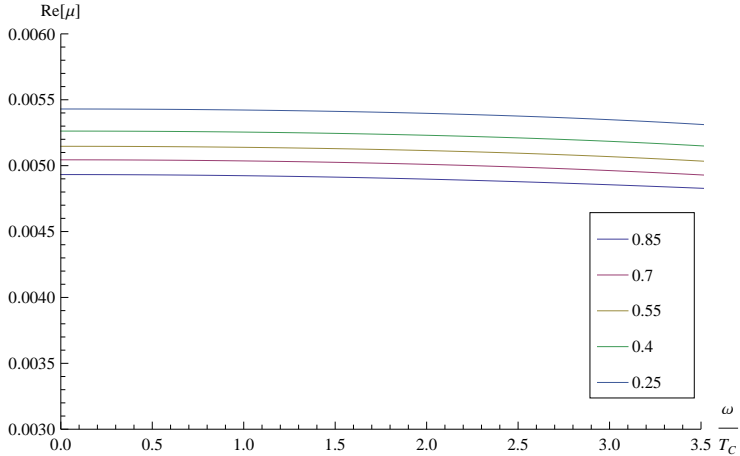}\\
  \caption{The real part of the effective magnetic permeability as a function of frequency with various temperatures.}\label{permeabilityreal}
\end{figure}

\begin{figure}

  \includegraphics[width=10cm]{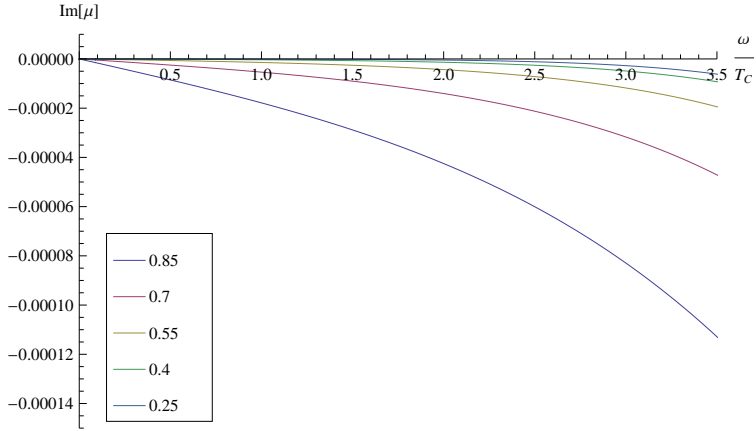}\\
  \caption{The imaginary part of the effective magnetic permeability as a function of frequency with various temperatures.}\label{permeabilityimaginary}
\end{figure}

Next we plot the real and imaginary parts of the effective magnetic
permeability in Fig.\ref{permeabilityreal} and
Fig.\ref{permeabilityimaginary} respectively. The real part of the
effective magnetic permeability still displays the similar behavior
at low frequencies as that in \cite{AFMP}, while the imaginary part
shows a totally different behavior, i.e., there is no blow-up at
$\omega=0$.

\begin{figure}

  \includegraphics[width=10cm]{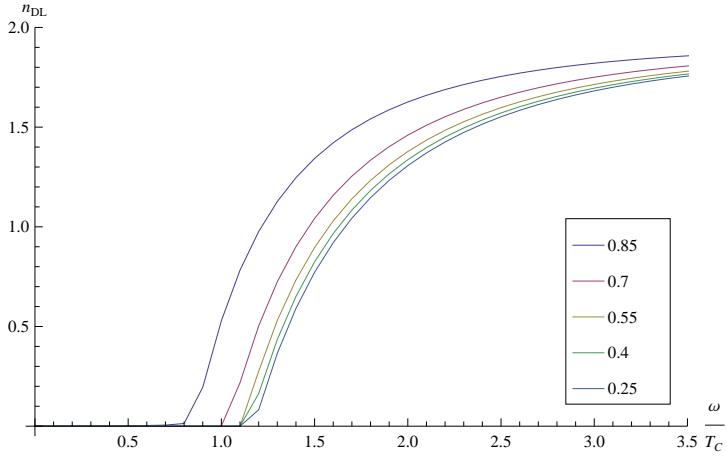}\\
  \caption{The Depine-Lakhtakia index as a function of frequency with various temperatures.}\label{dlindex}
\end{figure}

Such a difference eventually leads to the conclusion that at low
frequencies the negative phase velocity does not show up in the
holographic superconductors at least within the accuracy of our
numerics, which is illustrated by the non-negative Depine-Lakhtakia
index in Fig.\ref{dlindex}. In hindsight, the reason may arise in
the fact that the bulk background is essentially kind of neutral
black hole although the electromagnetic field has an influence on
the fluctuation equation (\ref{eomf}) indirectly through the
condensate.
\begin{figure}

  \includegraphics[width=12cm]{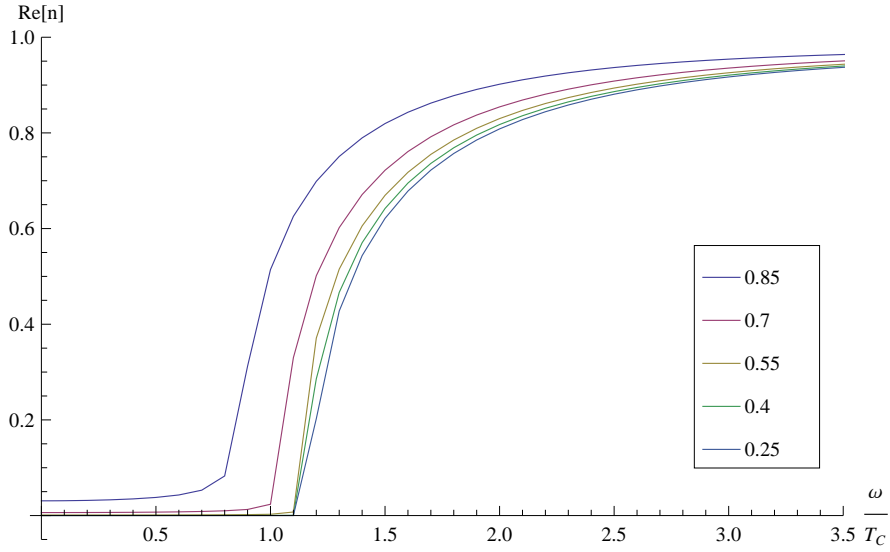}\\
  \caption{The real part of the refractive index as a function of frequency with various temperatures.}\label{realindex}
\end{figure}
\begin{figure}

  \includegraphics[width=12cm]{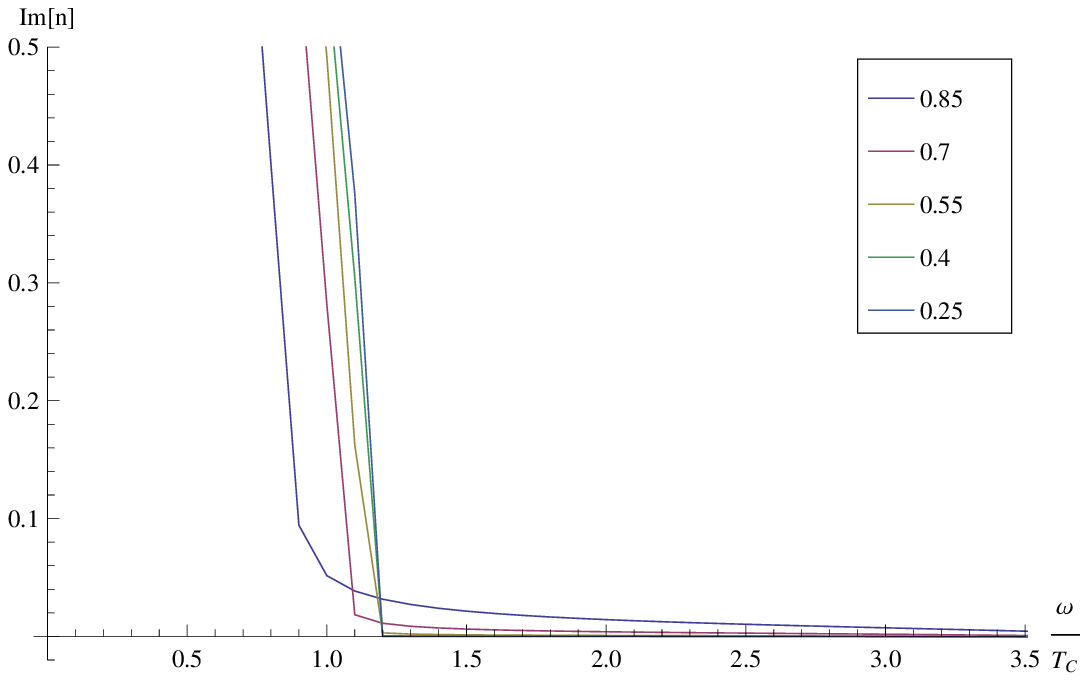}\\
  \caption{The imaginary part of the refractive as a function of frequency with various temperatures.}\label{imaginaryindex}
\end{figure}

At last, we would like to plot the real and imaginary parts of the
refractive index in Fig.\ref{realindex} and Fig.\ref{imaginaryindex}
respectively. Note that the ratio of the imaginary part of the
refractive index to its real part describes the ratio between
dissipation and propagation. We see that the there exists a
characteristic frequency, below which light can not propagate in our
holographic superconductors while above which light can propagate
without dissipation.
\section{Conclusions}
We have carried out the numerical analysis of the optical properties
of the s-wave holographic superconductors in the probe limit. In
particular, we have calculated the electric permittivity, effective
magnetic permeability, refractive index, and Depine-Lakhtakia index
by the holographic duality, which is generically difficult to be
achieved by other approaches. As a result, the electric permittivity
and effective magnetic permeability conspire to make our holographic
superconductors robust against a negative phase velocity at low
frequencies.

 We conclude with some generalizations of our work in various
directions. Firstly, although the cases for other scalar masses are
expected to demonstrate the qualitatively similar properties as the
massless case, it is interesting to investigate how the specific
optical properties of holographic superconductors depend
quantitatively on the mass $m$. Secondly, it is natural to extend
our work to the 2+1 dimensional holographic
superconductors\cite{GZ}. More importantly, as the superconductors
are believed to have the potential to support low losses, which is
critical for many applications like super-resolution imaging, it is
rewarding to search for the negative phase velocity in other
situations. In particular, it is worthwhile to go beyond the probe
limit to explore the full behavior of the theory, where the negative
refractive index is expected to appear at low frequencies because
the background is essentially kind of charged black hole. In
addition, as suggested by the early experimental implementation of
negative refractive index in the
ferromagnet-superconductors\cite{PLPD}, it is highly possible to see
the negative refractive show up when the external magnetic field is
added to our holographic superconductors. Furthermore, it is also
intriguing to extend our analysis to both the p-wave and d-wave
holographic superconductors, where some new features should come out
due to the nonisotropy of the dual media. Finally, it is also
tempted to generalize our work to some non-minimal models of
holographic superconductors\cite{FGR,AR,Pan,PW,LS,CNZ}.
\section*{Acknowledgements}
XG would like to thank Yan Liu, Rongxin Miao, Zhangyu Nie, and Yang
Zhou for useful discussions. HZ is much indebted to Antonio Amariti
and Davide Forcella for illuminating correspondence during the whole
project. He is also grateful to Bom Soo Kim, Elias Kiritsis, Matthew
Lippert, and Rene Meyer for helpful discussions. In addition, it is
a great pleasure to thank Ioannis Iatrakis and Qiyuan Pan for help
with the numerical calculation. XG was supported in part by the NSFC
under grant No. 10821504. HZ was partially supported by a European
Union grant FP7-REGPOT-2008-1-CreteHEPCosmo-228644.

\end{document}